\documentclass[twocolumn,epjc3]{svjour3} 
\input{preamble.tex}

\begin{document}
\title{Detection of Astrophysical Tau Neutrino Candidates in IceCube}
\journalname{Eur. Phys. J. C}
%
%\title{IceCube Author List for EPJC 20201016}
\onecolumn
\author{R. Abbasi\thanksref{loyola}
\and M. Ackermann\thanksref{zeuthen}
\and J. Adams\thanksref{christchurch}
\and J. A. Aguilar\thanksref{brusselslibre}
\and M. Ahlers\thanksref{copenhagen}
\and M. Ahrens\thanksref{stockholmokc}
\and C. Alispach\thanksref{geneva}
\and A. A. Alves Jr.\thanksref{karlsruhe}
\and N. M. Amin\thanksref{bartol}
% ...
}
\authorrunning{IceCube Collaboration}
\thankstext{a}{also at Universit{\`a} di Padova, I-35131 Padova, Italy}
%\thankstext{b}{also at National Research Nuclear University, Moscow Engineering Physics Institute (MEPhI), Moscow 115409, Russia}
\thankstext{c}{also at Earthquake Research Institute, University of Tokyo, Bunkyo, Tokyo 113-0032, Japan}
\institute{
% ...
Universit{\'e} Libre de Bruxelles, Science Faculty CP230, B-1050 Brussels, Belgium \label{brusselslibre}
% ...
\and Department of Physics, Loyola University Chicago, Chicago, IL 60660, USA \label{loyola}
\and Dept. of Physics and Astronomy, University of Canterbury, Private Bag 4800, Christchurch, New Zealand \label{christchurch}
% ...
\and Niels Bohr Institute, University of Copenhagen, DK-2100 Copenhagen, Denmark \label{copenhagen}
% ...
\and D{\'e}partement de physique nucl{\'e}aire et corpusculaire, Universit{\'e} de Gen{\`e}ve, CH-1211 Gen{\`e}ve, Switzerland \label{geneva}
% ...
\and Karlsruhe Institute of Technology, Institute for Astroparticle Physics, D-76021 Karlsruhe, Germany  \label{karlsruhe}
% ...
\and Bartol Research Institute and Dept. of Physics and Astronomy, University of Delaware, Newark, DE 19716, USA \label{bartol}
% ...
\and Oskar Klein Centre and Dept. of Physics, Stockholm University, SE-10691 Stockholm, Sweden \label{stockholmokc}
% ...
\and DESY, D-15738 Zeuthen, Germany \label{zeuthen}
}
\date{Received: date / Accepted: date}
 
\maketitle
\twocolumn
%\title{Detection of Astrophysical Tau Neutrinos in IceCube}
%\maketitle
\begin{abstract}
High-energy tau neutrinos are rarely produced in atmospheric cosmic-ray showers or at cosmic particle accelerators, but are expected to emerge during neutrino propagation over cosmic distances due to flavor mixing. When high energy tau neutrinos interact inside the IceCube detector, two spatially separated energy depositions may be resolved, the first from the charged current interaction and the second from the tau lepton decay. We report a novel analysis of 7.5 years of IceCube data that identifies two candidate tau neutrinos among the 60 ``High-Energy Starting Events'' (HESE) collected during that period. The HESE sample offers high purity, all-sky sensitivity, and distinct observational signatures for each neutrino flavor, enabling a new measurement of the flavor composition. The measured astrophysical neutrino flavor composition is consistent with expectations, and an astrophysical tau neutrino flux is indicated at 2.8$\sigma$ significance.
%
%We present the results of a search for astrophysical tau neutrinos in 7.5 years of IceCube's High-Energy Starting Event data. At high energies, two energy depositions stemming from the tau neutrino charged-current interaction and subsequent tau lepton decay may be resolved. We report the first detection of two such events, with probabilities of $\sim 76\%$ and $\sim 98\%$ of being produced by astrophysical tau neutrinos. The resultant astrophysical neutrino flavor measurement is consistent with expectations, and an astrophysical tau neutrino flux is indicated at 2.8$\sigma$ significance.
%
%Of the three neutrino flavors, the third-generation ``tau'’ neutrinos are of particular interest to astrophysicists.  Since they are rarely produced in atmospheric cosmic ray showers, each candidate tau neutrino serves as a window onto its origins in a cosmic particle accelerator, and as a probe of the fundamental physics affecting its propagation to Earth. Here we report a novel analysis of 7.5 years of IceCube data that identifies two candidate tau neutrinos among the 60 ``High-Energy Starting Events'' (HESE) collected during that period. This enhanced HESE sample offers high purity, all-sky sensitivity, and distinct observational signatures for both muon and tau neutrino flavors, enabling a new measurement of the HESE flavor distribution. We find this distribution to be broadly consistent with equipartition, as anticipated by most common scenarios, with a nonzero astrophysical tau neutrino flux indicated at 2.8$\sigma$ confidence.
%
\end{abstract}

\section{Introduction}
\label{sec:introduction}
The discovery of a diffuse flux of astrophysical neutrinos, using High-Energy Starting Events (HESE) observed by IceCube~\cite{IceCube} opened the possibility to study the Universe's most powerful cosmic accelerators~\cite{HESE2,HESE3}. HESE is an all-flavor, all-sky selection of events of predominantly astrophysical origin, with an analysis region above $60$~TeV in deposited electromagnetic-equivalent energy in the detector. Tau neutrinos are expected to be produced only in tiny fractions at neutrino sources, but emerge due to neutrino oscillations over cosmic baselines~\cite{Learned}. For neutrinos from distant sources, the probability of a neutrino created with flavor $\nu_{\alpha}$ to reach the detector as $\nu_{\beta}$ is $\langle P_{\nu_{\alpha} \to \nu_{\beta}} \rangle = \sum_i \vert U_{\alpha i}\vert^2 \vert U_{\beta i} \vert ^2 $ \cite{PNonconservation,MNS}.
Thus, the neutrino flavor composition at Earth depends on the neutrino mixing matrix elements, $U_{\alpha i}$, and the source flavor composition.
For neutrinos from the decay of charged pions produced in hadronic interactions, with a source flavor composition of $\nuratio=1/3:2/3:0$, we expect $\nuratio=0.30 : 0.36 : 0.34$ at Earth~(using the oscillation parameters from \cite{NuFit41}), i.e., very close to equipartition ($1/3:1/3:1/3$). However, the environment at the neutrino production sites may influence the flavor composition, due to cooling or interactions of the charged particles produced in the hadronic interactions~\cite{Mudamping,mubeam-nu,neutron-nu,charm-nu}. Therefore, the flavor composition of astrophysical neutrinos is a powerful probe of the environments of cosmic accelerators and can help constrain the source populations contributing to the observed neutrino flux. The neutrino flavor composition on Earth is also a sensitive probe of physics beyond the Standard Model (BSM) affecting neutrino propagation and modifying the flavor composition~\cite{Barenboim:decay,Keranen:sterile,Carlos:EffOp,MauricioWalter,Rasmus}; see~\cite{IceCube:2021tdn} for BSM-constraints derived using the HESE selection.

Atmospheric neutrinos are a background to astrophysical neutrino searches. 
As atmospheric neutrinos are accompanied by muons born in the same cosmic-ray-induced shower, their contribution to a sample can be suppressed by muon-rejecting event selection criteria, e.g.\ by using the outer parts of the detector as a vetoing region. This effect, called atmospheric neutrino self-veto \cite{selfveto}, is used in HESE \cite{HESE7}. Conventional atmospheric neutrinos are $\nu_{e}$ and $\numu$ from the decay of $\pi^{\pm}$ and $K^{0,\pm}$ produced in the atmosphere by cosmic-ray interactions. At energies above $\sim 100$~TeV, the atmospheric flux is expected to be increasingly dominated by the prompt component, originating from the decays of charmed hadrons (e.g. \cite{BERSS}). Tau neutrinos, produced from rare decays of $D_s$ and $D^{0,\pm}$, contribute only up to 5\% to the yet unobserved prompt atmospheric neutrino component \cite{ERS,AFTG}. This makes the observation of high-energy tau neutrinos a smoking-gun signature of cosmic neutrinos, but so far, none have been identified \cite{IC22,DXu,ICRCMarcel}.
Previous flavor studies only separated the charged-current $\nu_\mu$ contribution from other flavors, leading to a significant degeneracy between the $\nu_e$ and $\nu_{\tau}$ flavors \cite{Flavor1:Mena,APJ15,Spencer1}. Here, we present a new flavor composition measurement of astrophysical neutrinos with direct sensitivity to each of the neutrino flavors, performed on the HESE sample. A detailed description of the characteristics of the HESE sample and spectral fits to a diffuse astrophysical neutrino spectrum assuming flavor equipartition, as well as a detailed description of systematic uncertainties and their treatment are provided in~\cite{HESE7}.
There, the astrophysical neutrino spectrum was fit as a single power law,
\begin{linenomath*}
\begin{equation}
    \frac{\dd \Phi_{\nu}}{\dd E} = \phi_{\nu} \cdot  \left ( \frac{E}{E_0} \right )^{\gammaa},
	\label{eq:Phitot}
\end{equation}
\end{linenomath*}
where $\phi_{\nu}$ is the all-flavor $\nu+\overline{\nu}$ flux at $E_0=100$~TeV and $\gammaa$ is the spectral index.
Their best fit values are 
$\phi_{\nu}=6.4^{+1.5}_{-1.6} \, \times \, 10^{-18}  \mathrm{\, GeV^{-1} \, cm}^{-2} \mathrm{\, s}^{-1} \mathrm{\, sr}^{-1}$, and 
$\gammaa=-2.87^{+0.21}_{-0.19}$.
The sample and associated results have been made available publicly through a dedicated data release~\cite{HESE7data}.

This manuscript is structured as follows: Section~\ref{sec:signatures} describes the signatures of neutrino interactions detected in IceCube and how they map to neutrino flavors; Section~\ref{sec:classification} illustrates the selection and classification of the detected events according to these various signatures; Section~\ref{sec:classresults} summarizes the outcome of the classification and the characteristics of the found $\nu_{\tau}$ candidates; in Section~\ref{sec:flavorcomp} the flavor composition constraints from this analysis are derived. 

\section{Neutrino Signatures in the Detector}
\label{sec:signatures}
In IceCube, neutrinos are detected by collecting the Cherenkov light emitted by charged secondary particles created in neutrino interactions. All neutral-current (NC) interactions produce showers of hadrons and are indistinguishable between flavors. 
In a charged-current (CC) interaction, the neutrino flavor can be inferred from the distinct Cherenkov light pattern produced by each flavor of charged lepton. 
Light depositions from a muon traversing the detector are called \textit{tracks} and stem from $\nu_{\mu}$ CC interactions, atmospheric muons, and $\nu_{\tau}$ CC interactions where the tau decays to a muon (17\% branching ratio). \textit{Single cascades} consist of energy depositions at a single vertex and are produced by $\nu_{e}$ CC and NC interactions of all flavors. At PeV energies, both tracks and single cascades can also emerge from the decays of W-bosons produced in resonant neutrino-electron scattering~\cite{IceCube:2021rpz}. \textit{Double cascades} are two energy depositions connected by a track of comparatively low light emission. They are produced by $\nutau$ CC interactions where the first cascade originates from the hadronic interaction of the $\nutau$ producing a tau, and the second cascade stems from the tau decaying to a hadronic or electromagnetic cascade (83\% branching ratio) \cite{Learned}. Due to their short livetime, taus have a short decay length of $\langle L_{\tau} \rangle \sim 50$~m $\cdot \, E_{\tau}$~/~PeV, where $ E_{\tau}$ is the tau energy. This makes the distinction between single and double cascades challenging in IceCube, where the mean horizontal distance between light sensors, called Digital Optical Modules (DOMs), is 125~m. The HESE analysis defines a lower threshold on the deposited electromagnetic-equivalent energy in the detector of events, $\etot$, of 60~TeV (see below). Above this threshold it is possible to identify some of the $\nu_{\tau}$ events as double cascades, if $L_{\tau} \gtrsim 10$~m, breaking the degeneracy between $\nu_e$ and $\nu_{\tau}$ flavors present at lower energies\footnote{It may, however, be possible to distinguish $\nue$ and $\nutau$ events on a statistical basis at lower energies, e.g., using the method proposed in~\cite{Beacom:echo}.}. 

\section{Event Selection and Classification}
\label{sec:classification}
Using the HESE selection, we have performed a new analysis of the IceCube data that incorporates major improvements with respect to previous publications \cite{HESE2, HESE3} in our understanding of the detector and the modelling of atmospheric backgrounds.
The HESE selection is  described in \cite{HESE2}. To pass, an event has to (1) start inside of the outermost layer of DOMs making up the ``veto'' layer, and (2) deposit more than 6000 photoelectrons in the detector. Muons radiate away energy throughout their passage through the ice, with the amount of light deposited increasing with increasing muon energy. It is thus extremely unlikely for atmospheric muons to pass the HESE selection criteria. Due to the atmospheric self-veto (\cite{selfveto}, see also Section~\ref{sec:introduction}), accompanying muons also greatly reduce the number of downgoing atmospheric neutrinos present in the sample. To further enhance the fraction of astrophysical neutrinos in the sample, the analysis is restricted to events with a reconstructed total deposited energy $\etot$ above $60$~TeV.
Data collected between 2010 to 2017 using the original HESE selection \cite{HESE2}, with a total livetime of 2635 days, have been reprocessed using a new and improved detector calibration. An improved model of the optical properties of the South Pole ice sheet \cite{Chirkin:2013lpu}, critical to the reconstruction of event properties, has been incorporated into the simulation and reconstruction, and an updated calculation of the atmospheric neutrino self-veto \cite{APRSWY} is used. This new HESE sample has 60 events in the analysis region, i.e.\ with $\etot > 60$~TeV, and is described in detail in \cite{HESE7}. 

\begin{figure}[tb]
    \centering
    \includegraphics[width=80mm]{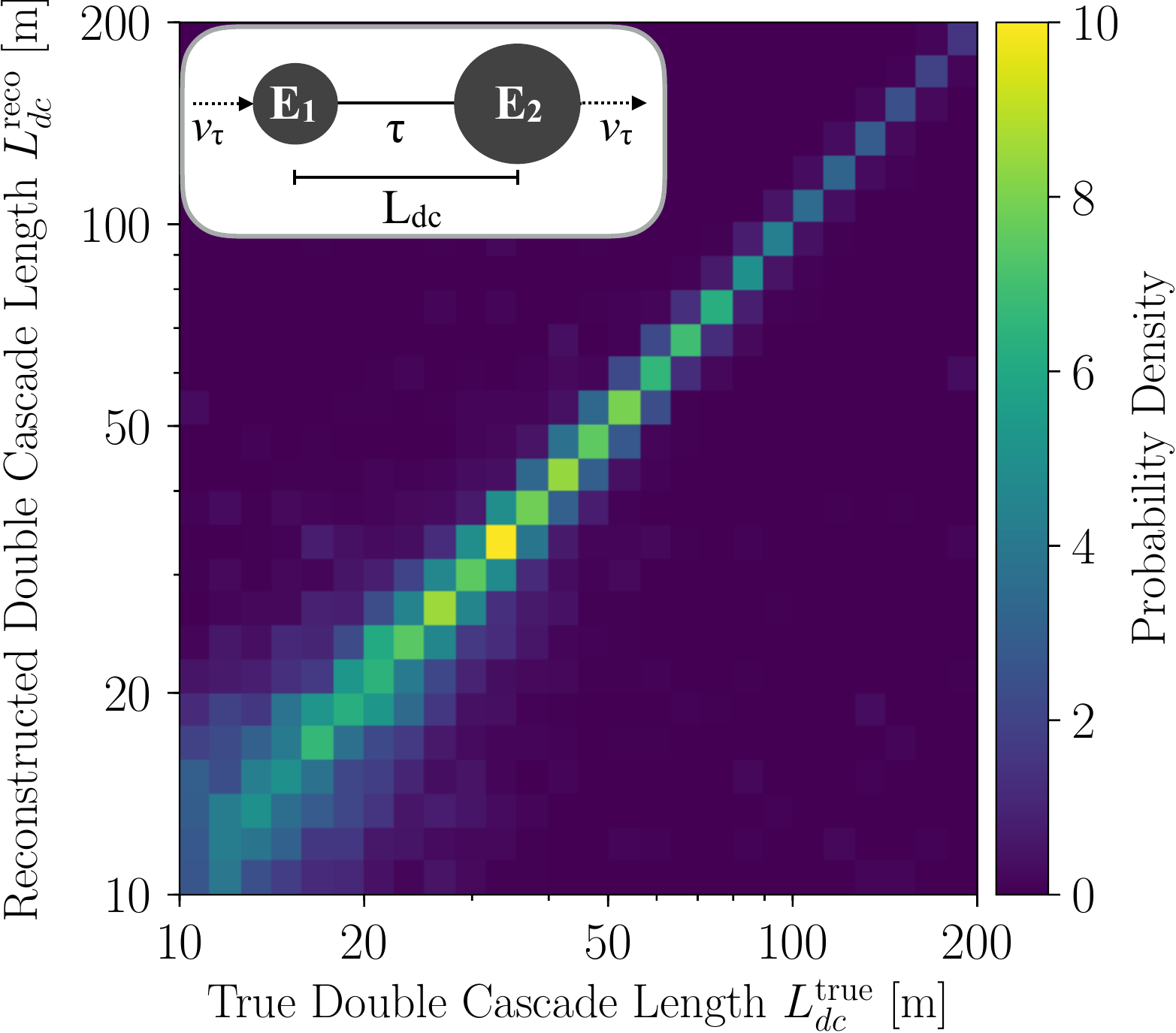}
    \caption{Length resolution for simulated tau neutrinos classified as double cascades for the best-fit spectrum \cite{HESE7}. With increasing length, the resolution improves while the event expectation drops. The inlay shows a schematic of a $\nutau$ CC interaction producing a double cascade with associated reconstruction parameters. }
    \label{fig:Tau_L_L}
\end{figure}
We use a classification algorithm developed on Monte Carlo (MC) simulated events and first applied to the six-year HESE sample \cite{HESE6, ICRCMarcel} to classify the 60 events as single cascades, double cascades, or tracks (ternary event classification). It was developed with the goal of achieving a high $\nu_{\tau}$ purity for the events assigned a double-cascade topology, while keeping misclassification fractions low for all topologies \cite{MarcelPhD}.
All events are reconstructed using maximum likelihood fits with different hypotheses: single cascade \cite{Ereco}, track \cite{SPE}, and double cascade  \cite{Ereco,Patrick}. For the fits, the timing and spatial information of the light collected in an event is used. The parameters of the double-cascade fit are (see also inset of Figure \ref{fig:Tau_L_L}): the energies of the interaction and decay cascade, $E_1$ and $E_2$ respectively; the spatial separation between them (called \textit{double-cascade length} $\ldc$ hereafter); the direction and vertex of the first cascade. The total energy $\etot$ of the event is the sum of all energy depositions obtained from a track energy unfolding; for double cascades this equals $E_1+E_2$.
The two cascades are assumed to be co-linear due to the large Lorentz boost.

A preselection removes events with a failed double-cascade fit from being further considered as double cascades. After preselection, three event properties are used to classify each event: \textit{double-cascade length}, \textit{energy asymmetry}, and \textit{energy confinement}. The double-cascade length is a proxy for the tau lepton’s decay length with an average resolution of $\sim$2~m over the entire analysis range at the best-fit spectrum with flavor equipartition~\cite{HESE7}. Figure~\ref{fig:Tau_L_L} shows the reconstructed double cascade length as a function of the true double cascade length; the length resolution improves with increasing length as the cascades get better separated. The energy asymmetry is defined as $A_E=(E_1-E_2)/(E_1+E_2)$. It can take values $-1 \leq A_E \leq 1$, with the boundary values corresponding to single cascades. The energy confinement is defined as $E_C=(E_{C1}+E_{C2})/E_{\mathrm{tot}}$, where $E_{Ci}$ are the energy depositions within 40~m of the $i$-th cascade vertex position. For the purpose of this calculation the vertices of the two cascades are taken directly from the double-cascade reconstruction, while the energy depositions are obtained through a track energy unfolding algorithm~\cite{Ereco}, and thus the confinement can take values $0 < E_{C} < 1$. Thus, for double cascades separated by $\lesssim 80$~m  the relation $\etot = E_1+E_2 = E_{C1}+E_{C2}$ holds. Events passing the requirements shown in the second column of Table \ref{tab:classification} are classified as double cascades. 

True single cascades typically have a small reconstructed double-cascade length and a large, positive energy asymmetry. True tracks typically have energy depositions all along their tracks, leading to low energy confinement values. True double cascades have $E_C$ values very close to $1$  even for separation lengths in excess of $80$~m, due to the low relative brightness of the tau. By choosing a conservative requirement of $E_C > 0.99$ for double cascades, the performed analysis does not lose sensitivity towards higher-energy $\nutau$ producing longer-lived $\tau$ leptons. True double cascades show a flat distribution in $A_E$ with a resolution of $\sim 0.1$ at negative values of $A_E$ and worsening towards positive values.
Their double-cascade length is correlated to their total deposited energy and follows the exponentially falling distribution seen in the energy spectrum.
Events failing the double-cascade requirements are classified according to the procedure shown in the last column of Table \ref{tab:classification}.
\begin{table}[tb]
\centering
\begin{tabular}{c | cc}
\multirow{2}{*}{\makecell[c]{Observable}} &  Requirement for & Classification if \\ & double cascade & requirement failed \\[-0.8em] \\
\hline \hline
\\[-0.8em]
$\etot$  & $\geq 60$~TeV & Event rejected \\ [-0.8em] \\
\hline
\\[-0.8em]
\multirow{2}{*}{Preselection} & \multirow{2}{*}{passed} & Single cascade / track \\
             &        & (depending on likelihood) \\[-0.8em] \\ 
\hline \\[-0.8em]
$\ldc$ & $\geq10$~m & Single cascade \\ [-0.8em] \\ 
\hline \\[-0.8em]
$E_C$ & $\geq 0.99$ & Track \\ [-0.8em] \\
\hline \\[-0.8em]
$A_E$ & $\in [-0.98, 0.30]$ & Single cascade \\[-0.8em] \\
\hline
\end{tabular}
\caption[Ternary topological classification chain]{Steps for the ternary topological classification in order of precedence. For events failing the ``preselection'', the likelihoods of the track and single-cascade fits are compared, and the topology with the higher likelihood fit is chosen.}
\label{tab:classification}     
\end{table}
Note that the requirement of $\ldc\geq 10$~m for double cascades leads to the majority of $\nu_{\tau}$ induced events to be classified as single cascades. At the best-fit spectrum with flavor equipartition \cite{HESE7}, we expect $\sim15$ $\nutau$ events, of which $\sim 12$ interact via the double cascade channel. But only $\sim 3\, (22.7\%)$ of those are expected to produce a tau that travels at least 10~m before decaying. $42.3 \%$ of simulated double cascades with tau decay lengths above 10~m pass the double cascade requirements in Table~\ref{tab:classification}. The total efficiency of the ternary topological classification chain for double cascades is $12.2\%$. $1.9\%$ of all $\nue$ and $\numu$ induced events are expected to be misclassified as double cascades.

Glacial ice at South pole flows at a rate of $\sim 10$~m per year. It has recently been observed~\cite{tc-14-2537-2020} that the optical properties of glacial ice at South Pole vary as a function of the direction with respect to the flow of the glacial ice. This ice anisotropy is one of the limiting factors on the selection of double cascades: directional distortions of Cherenkov light patterns can lead to a misclassification of single cascades as double cascades. See Appendix~C for details on the ice anisotropy treatment.
\section{Results of the Topological Classification}
\label{sec:classresults}
\begin{figure}[tb]
\centering
\includegraphics[width=84mm,clip]{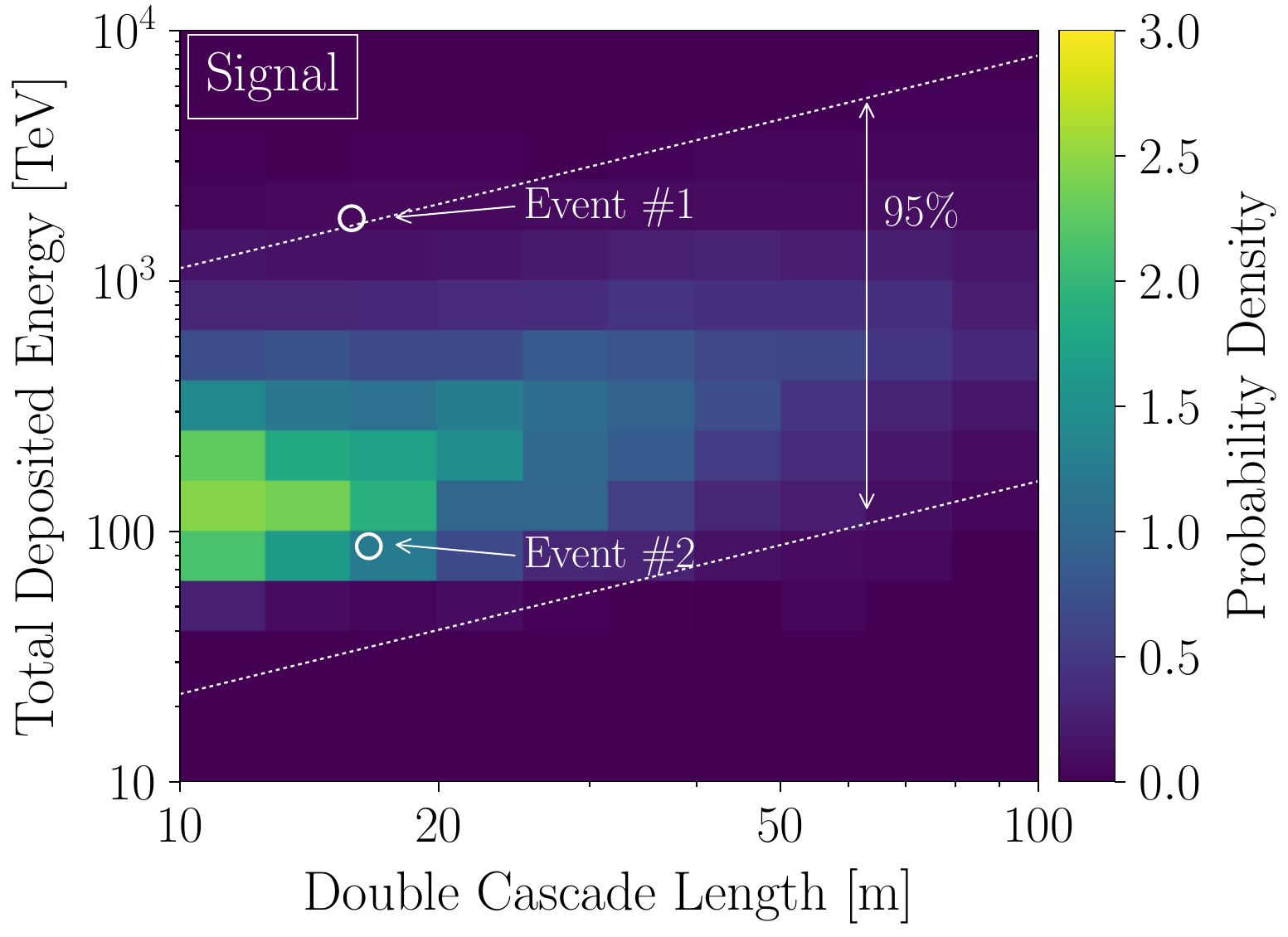}  \\
\includegraphics[width=84mm,clip]{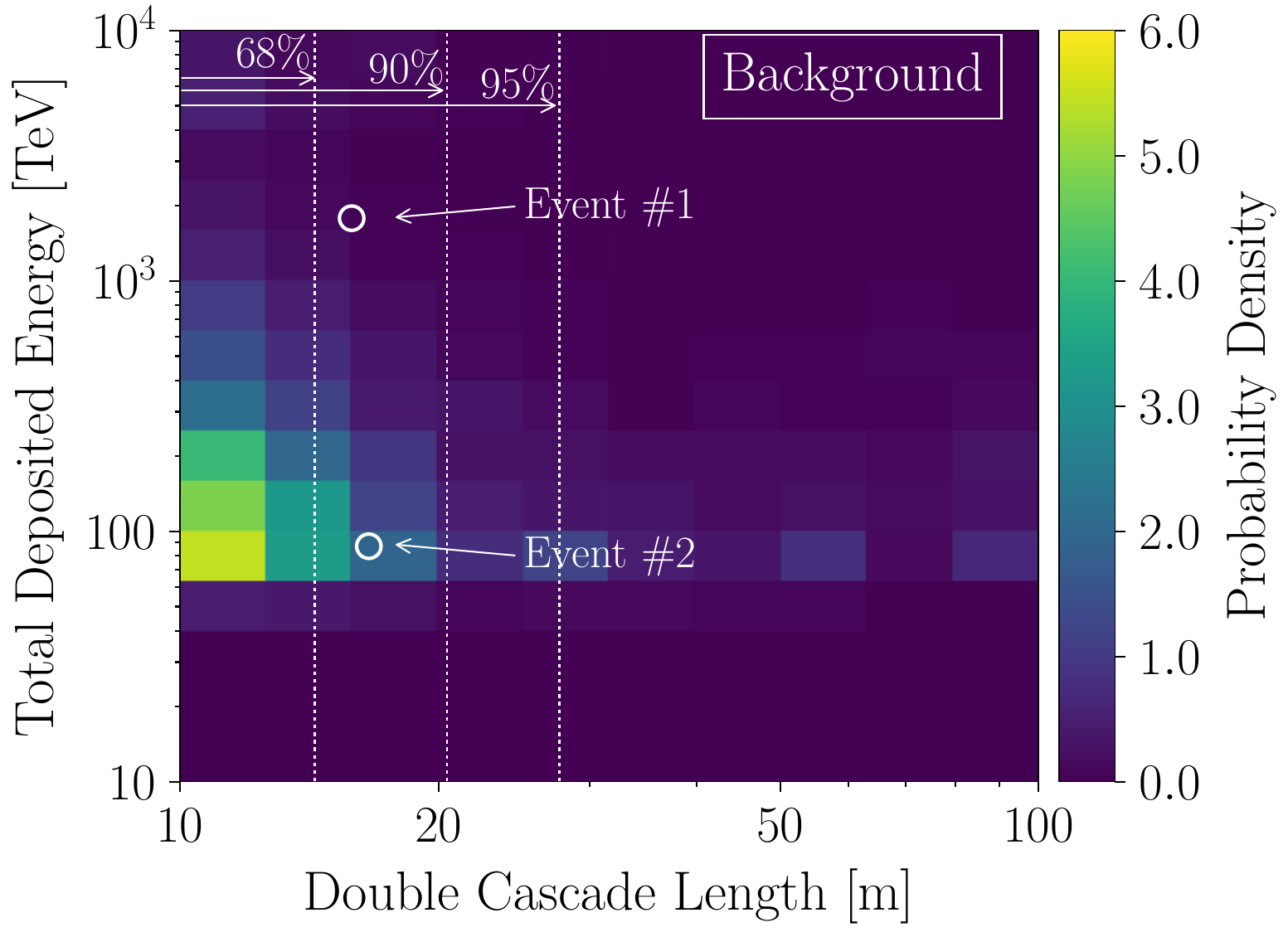}
\caption[Two-dimensional MC PDFs showing total reconstructed energy versus reconstructed double-cascade length for the double-cascade sample with data points]{Two-dimensional MC PDFs showing total reconstructed energy versus reconstructed double-cascade length for the double-cascade subsample with data points, using the best fit to the atmospheric and astrophysical components with the flavor composition of astrophysical neutrinos fixed to $1:1:1$ \cite{HESE7}. In the signal ($\nutau$-induced double-cascade events) histogram (\textit{top}), the region containing $95\%$ of the expected signal is indicated with white dotted lines. In the background (all remaining events) histogram (\textit{bottom}), the white vertical dotted lines mark the regions containing $68\%,\,90\%,$ and $95\%$ of the single-cascade induced background. In both histograms the two tau neutrino candidates are overlaid as white circles.}  \label{fig:PID}       
\end{figure}

\begin{table}[tb]
\centering
\begin{tabular}{l | ll}
  & Event \#1 & Event \#2  \\\hline
Year & 2012 & 2014 \\ 
Energy of 1st cascade & 1.2 PeV & 9 TeV \\
Energy of 2nd cascade & 0.6 PeV & 80 TeV \\
Energy Asymmetry & 0.29 & -0.80 \\
Double-cascade Length & 16 m & 17 m \\ \hline
\end{tabular}
\caption{Reconstructed properties of the two double cascades. Uncertainties are $\sim$~10\% for the deposited energy and $\sim$~2~m for the double-cascade length. }
 \label{tab:DC}
\end{table}
The 60 events are classified into 41 single cascades, 17 tracks, and 2 double cascades. 
These are the first double cascades in the signal region and the first astrophysical tau neutrino candidate events. 
The reconstructed properties of the double cascades are shown in Table~\ref{tab:DC}. As the average tau decay length scales with the tau energy $\langle L_{\tau} \rangle \sim 50$~m $\cdot \, E_{\tau}$~/~PeV, which depends on the energy of the incoming $\nutau$ as $\langle E_{\tau} \rangle \sim 0.7\cdot \, E_{\nutau}$, the double cascades length $\ldc$ scales with the total deposited energy $\etot$. 
Two-dimensional MC probability distribution functions (PDFs) of reconstructed total deposited energy $\etot$ versus reconstructed double cascade length $\ldc$ for signal and background contributions to events classified as double cascades are shown in Figure~\ref{fig:PID} with the data events overlaid as white circles. For $\nu_{\tau}$-induced double cascades (top panel), a clear correlation between $\etot$ and $\ldc$ can be seen. Background events (bottom panel) cluster at the thresholds in $\etot$ due to the falling spectrum and in $\ldc$ since single cascades typically have very small reconstructed $\ldc$. The regions containing 68\%, 90\%, and 95\% of true single cascades misclassified as double cascades are marked by vertical white lines, i.e.\ 68\% of the true single cascades misclassified as double cascades have $\ldc<14.4$~m, while 90\% have $\ldc<20.4$~m. The tilted white lines show the region within which 95\% of the signal are contained. Few events are expected in the parameter space of event \#1, while there are contributions expected from both signal and background in the parameter space of event \#2. For single cascades and tracks, the properties \textit{total deposited energy}, $\etot$, and \textit{cosine of the zenith angle} in detector coordinates, $ \cos(\theta_z)$, are used to distinguish atmospheric and astrophysical contributions. The PDFs shown in Figure \ref{fig:PID} and the corresponding PDFs for single cascades and tracks described above are used in the all-flavor analyses presented in \cite{HESE7}. 

\subsection{Double-Cascade Event Characteristics}
\begin{figure}[htb]
\centering
\includegraphics[width=84mm,clip]{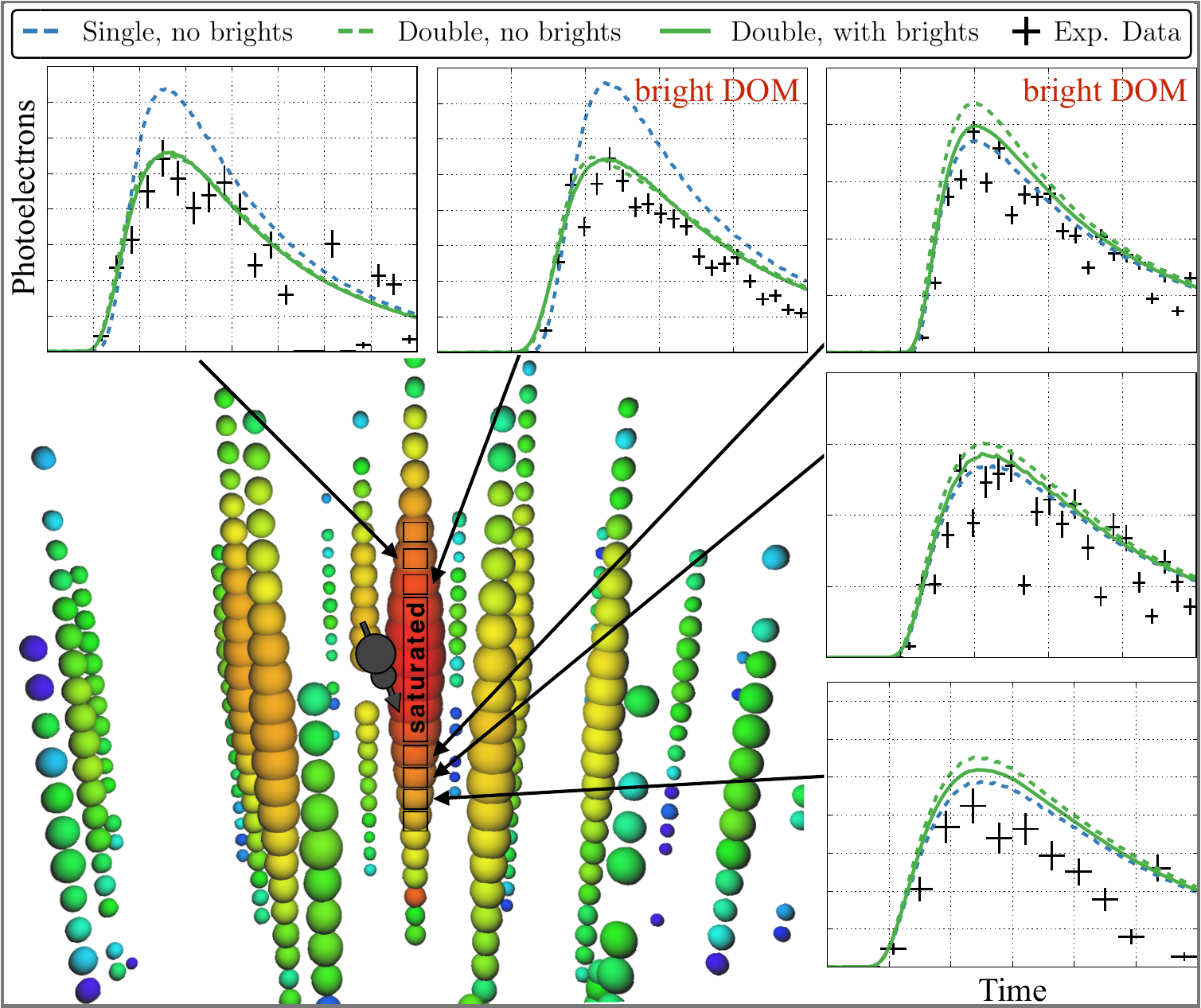}
\caption{Double-cascade event~\#1 (2012). The reconstructed double-cascade vertex positions are indicated as grey circles, the direction indicated with a grey arrow. The size of the circles illustrates the relative deposited energy, the color encodes relative time (from red to blue). Bright and saturated DOMs are excluded from this analysis.} 
 \label{fig:panoptBB}
\end{figure} 
An event view of event~\#1, observed in 2012 and nicknamed ``Big Bird'' \cite{HESE3}, is shown in Figure \ref{fig:panoptBB}. For several DOMs, the photon counts as a function of time are displayed alongside the predicted photon count distributions for single- and double-cascade hypotheses. The double-cascade hypothesis fits the observed data better than the single-cascade hypothesis.
However, this event has several saturated and bright DOMs that were excluded from the analysis, a standard procedure for high-energy IceCube analyses \cite{CCscd,TGM1}. A DOM is called \textit{saturated} if the signal in the PMT exceeds the dynamic range of the readout electronics. A DOM is called \textit{bright} if it has collected ten times more light than the average DOM for an event. Only statistical uncertainties on photon count rates are included in the likelihoods of the reconstruction algorithms \cite{Ereco,SPE,Patrick}. At the highest observed energies, bright DOM signals have very small statistic uncertainties and  can therefore lead to misreconstructions due to the lack of proper systematic uncertainty terms in the likelihood. For comparison of predicted photon counts for each hypothesis, the \textit{bright} DOMs are displayed in Figure~\ref{fig:panoptBB}.

\begin{figure}[t]
\centering
\includegraphics[width=84mm,clip]{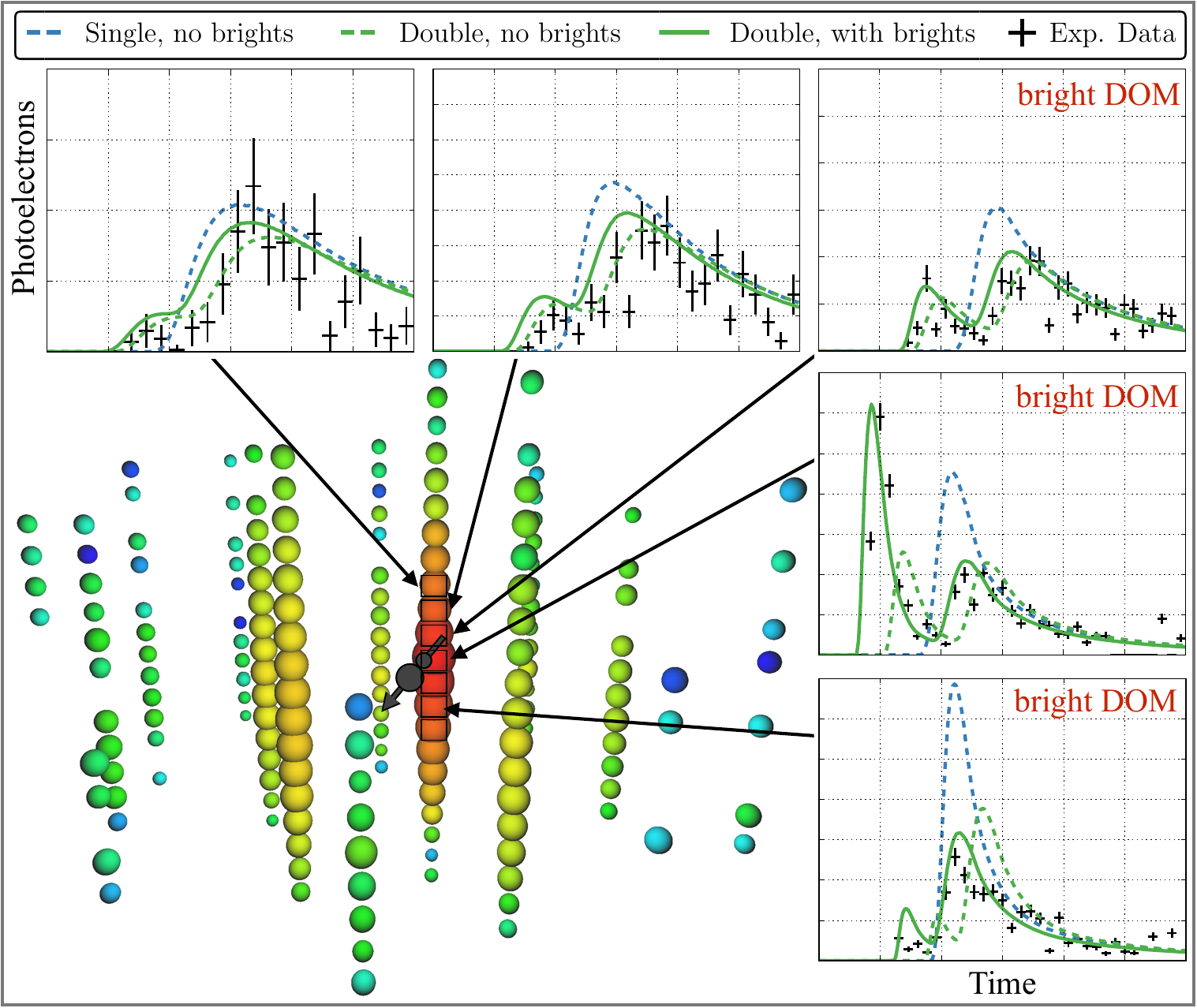}
\caption{Double-cascade event~\#2 (2014). The reconstructed double-cascade vertex positions are indicated as grey circles, the direction indicated with a grey arrow. The size of the circles illustrates the relative deposited energy, the color encodes relative time (from red to blue). Bright DOMs are excluded from this analysis.} 
 \label{fig:panopt} 
 \end{figure}

An event view of event~\#2, observed in 2014 and nicknamed ``Double Double,'' is shown in Figure~\ref{fig:panopt}. The two vertices of the cascades cannot be spatially resolved by eye, highlighting the need for the algorithmic topological classification employed in this work. Analogous to Figure~\ref{fig:panoptBB}, collected photon counts as a function of time are displayed together with the predicted photon count distributions for single- and double-cascade hypotheses. The predicted photon count PDFs differ remarkably between the single- and double-cascade hypothesis, with the single-cascade hypothesis disfavored. 
Data from DOMs labeled as \textit{bright} were excluded from the analysis, but are used for the comparison of predicted photon count PDFs in Figure~\ref{fig:panopt}. 
\begin{figure}[h!]
\centering
\includegraphics[width=85mm,clip]{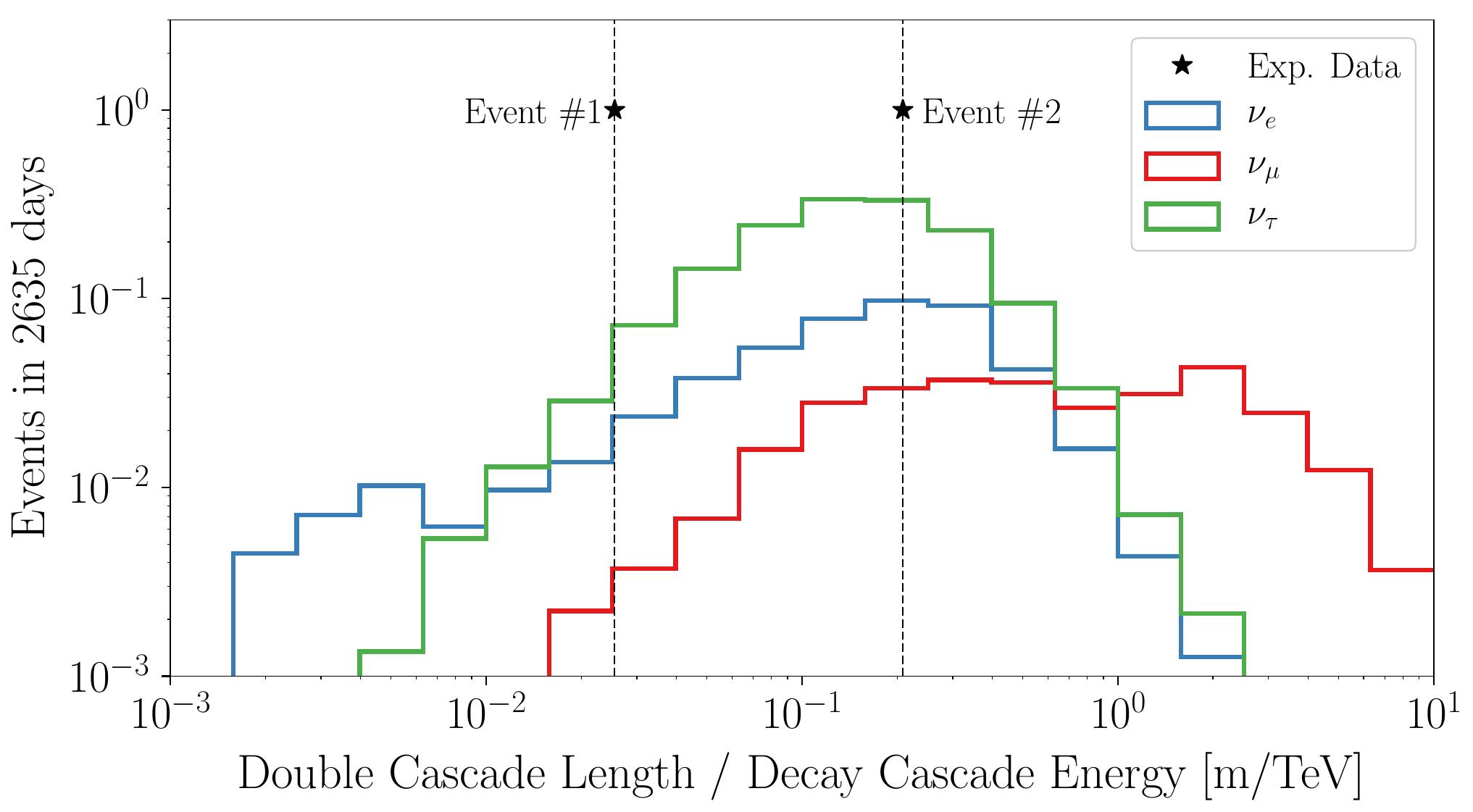} \\
\includegraphics[width=85mm,clip]{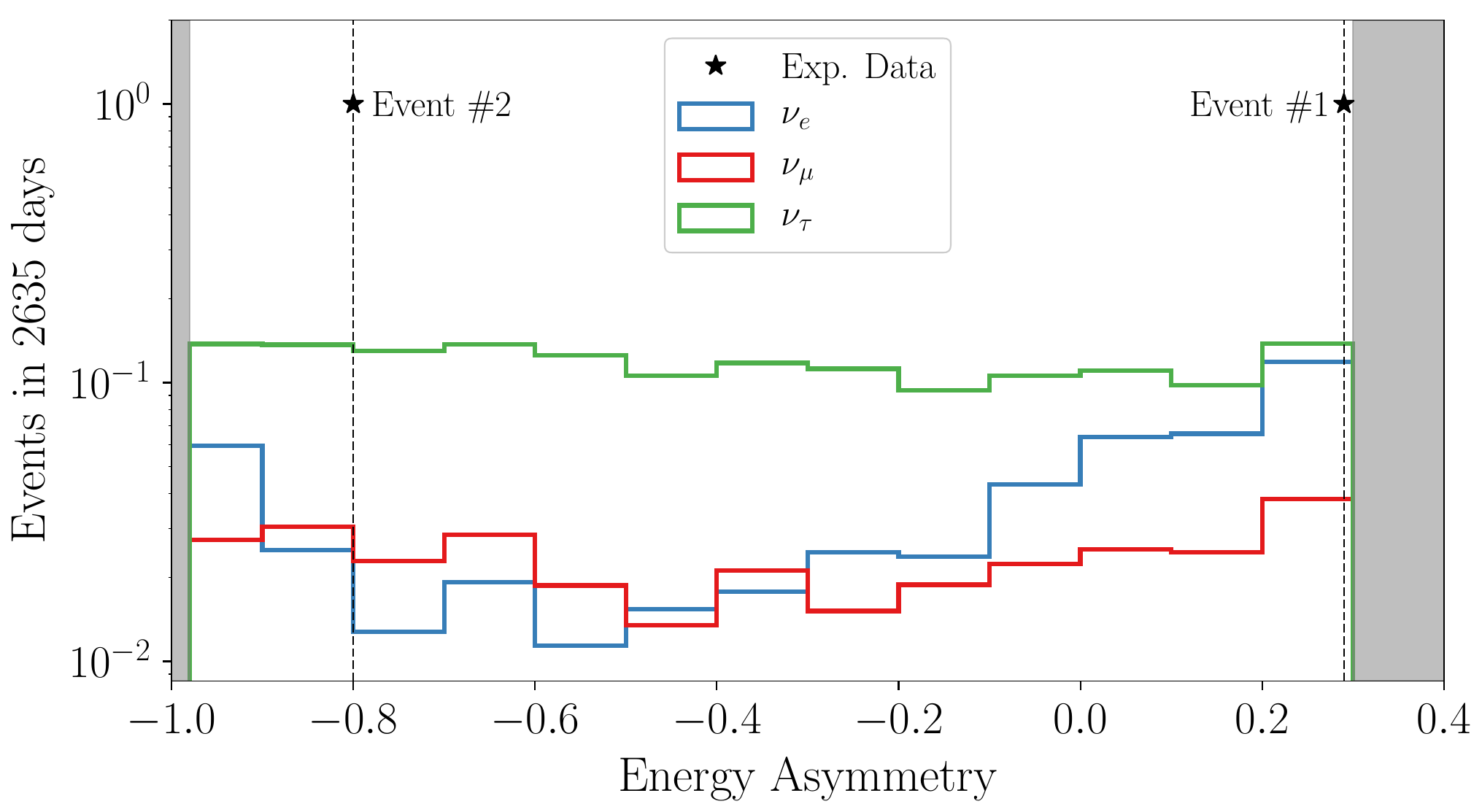} 
\caption{
Distribution of the ratio of double-cascade length to reconstructed decay-cascade energy (\textit{top}), and of the reconstructed energy asymmetry (\textit{bottom}) in the double-cascade subsample split by flavor content for the best-fit astrophysical and atmospheric spectra assuming flavor equipartition \cite{HESE7}. The values of the two double cascades are shown. Regions outside of the energy asymmetry values required for double cascades are marked in grey. }
 \label{fig:EL_EA}       
\end{figure}

Figure~\ref{fig:EL_EA} shows the distribution of the ratio of the double-cascade length $\ldc$ to reconstructed decay-cascade energy $E_2$ (top panel) and the energy asymmetry $A_E$ (bottom panel) of simulated events and data for the best-fit spectrum given in \cite{HESE7}. The distributions were not part of the topological classification chain. While the correlation between $\ldc$ and $\etot$ is clear on average, there are large fluctuations in energy transfer from parent to daughter particle. Therefore, on the per-event basis, the more direct correlation between the double-cascade length $\ldc$ and the decay-cascade energy $E_2$ proves more informative. 
Event~\#1 has a length-to-energy ratio in a region where the $\nutau$ contribution is larger than the background contribution, but outside of 90\% of the simulated $\nutau$-induced double cascades. Its high energy asymmetry is in a region with a background expectation which is on the order of the signal expectation. Event~\#2 has a length-to-energy ratio at the peak of the distribution for $\nutau$-induced double cascades and an energy asymmetry value in a highly signal-dominated region. None of the classified double cascades are in a phase space greatly affected by the ice anisotropy.
\subsection{Tau Neutrino Probability Assessment}
\label{sec:tauness}
\begin{table}[]
\centering
\begin{tabular}{l | cc}
Variable & Event \#1 & Event \#2  \\\hline
Primary Energy  & $>1.5$~PeV & $>65$~TeV \\
Visible Energy & 1 - 3~PeV & 60 - 300~TeV \\ 
Vertex, $r-r_{\mathrm{evt}}$ & 50~m & 50~m \\
Vertex, $z-z_{\mathrm{evt}}$ &  $\pm25$~m &  $\pm25$~m \\
Azimuth $\phi - \phi_{\mathrm{evt}}$ & $\pm 110 (40)\degree$  & $\pm 110\degree$ \\ 
Zenith $\theta - \theta_{\mathrm{evt}}$ & $\pm 35 (17)\degree$  & $\pm 35 \degree$ \\ \hline
\end{tabular}
\caption{Parameter space for resimulated events. The upper value of the primary energy depends on the interaction type, reflecting the spread of visible energy losses typical of that interaction. The visible energy is the energy transformed into light, it equals the total energy deposited in the detector for electromagnetic showers and is lower for hadronic showers and events with final-state muons or neutrinos. $r-r_{\mathrm{evt}}$ is the two-dimensional distance in the $x,y$-plane. The values in parentheses are for $\numu$ CC events.}
\label{tab:resim}     
\end{table}
To quantify the compatibility with a background hypothesis (i.e., not $\nutau$-induced) for the actual $\nutau$ candidate events observed, a targeted MC simulation for each event was performed, consisting of simulation of $\nue$, $\numu$, and $\nutau$ interactions. In addition, for ``Double Double,'' also atmospheric muons were simulated. However, none of the $1.2 \cdot 10^{10}$ generated muons passed the HESE veto undetected. See Table~\ref{tab:resim} for details on the restricted parameter space and Appendix A for a description of how this parameter space was chosen. Using targeted MC simulation for the analysis of exceptional events is a method often employed in IceCube \cite{IceCube:2013cdw,IceCube:2021rpz,HESE3,TXSarchival}. These new MC events were filtered and reconstructed in the same way as the initial MC and data events. In total, $\sim 2 \cdot 10^7$ ``Double-Double''-like events and $\sim 1 \cdot 10^6$ ``Big-Bird''-like events from the targeted simulation pass the HESE selection criteria. A breakdown of simulated event types and their fractions passing the HESE double cascade selection criteria can be found in Appendix~A. 

We define the tauness, $P_{\tau}$, as the posterior probability for each event to have originated from a $\nutau$ interaction, which can be obtained with Bayes' theorem: 
\begin{linenomath*}
\begin{equation}
    \begin{split}
        P(\nutau \mid \etaevt) &= \frac{P(\etaevt | \nutau) P(\nutau)}{P(\etaevt | \nutau) P(\nutau) + P(\etaevt | \nottau) P(\nottau)} \\
        & \approx \; \frac{N_{\nutau} P_{\nutau}(\etaevt)}{N_{\nutau} P_{\nutau}(\etaevt) + N_{\nottau} P_{\nottau}(\etaevt)} \\
        & \equiv \; P_{\tau}.
    \end{split}
\end{equation}
\end{linenomath*}
In the first line we have simply split the total probability of an event at the observed parameter space $\etaevt$ into its $\nutau$ and non-$\nutau$ (written $\nottau$) components in Bayes' theorem. In the second line we identify $P(\etaevt | \nutau)$ with the PDFs for $\nutau$, and express the prior probability $P(\nutau) = N_{\nutau}/(N_{\nutau}+N_{\nottau})$ as the fraction of expected $\nutau$ events evaluated at the observed parameter space of each event, $\etaevt$, obtaining the differential number of expected events $N_{\nutau} P_{\nutau}(\etaevt)$ (and analogous for the non-$\nutau$ components indicated as $\nottau$). 

For each tau neutrino candidate, the differential expected number of events at the point $\vec \eta_{\rm{evt}}$, $N_{\nutau} P_{\nutau}(\vec \eta_{\rm{evt}})$ and $N_{\cancel{\nutau}} P_{\cancel{\nutau}}(\vec \eta_{\rm{evt}})$ is approximated from the targeted simulation sets using a multidimensional kernel density estimator (KDE) with a gaussian kernel and the Regularization Of Derivative Expectation Operator (\textit{rodeo}) algorithm~\cite{rodeo}. 
The \textit{rodeo} algorithm provides an unbiased and computationally efficient way to find the optimal bandwidth in $d$ dimensions for a $d-$dimensional set of $n$ events. In the \textit{rodeo} the bandwidth is reduced as long a the derivative of the kernel density estimate with respect to its bandwidth is large compared to its variance. The obtained optimal bandwidth for each considered dimension balances the relevance of the variable with the sparsity of the dataset at the evaluated point. The eight dimensions used in evaluating the tauness include the six dimensions ($d=6$) of the restricted parameter space that the resimulation was carried out in: total deposited energy $\etot$, vertex position (\textit{x, y, z}) and direction ($\theta, \phi$). Further, a region of interest is defined in the parameters not restricted during resimulation but used in the double-cascade classification: double-cascade length $\ldc$ and energy asymmetry $A_E$~\cite{JSPhD}. The region of interest is obtained by slowly decreasing a two-dimensional box around the observed parameters as long as the statistical errors from the limited targeted MC stay below 10\%. This procedure was established using the produced MC in a sideband region.

Having defined  
$\etaevt = (\etot, x, y, z, \theta, \phi, \ldc, A_E)$, and approximating 
\begin{linenomath*}
\begin{equation}
N_{\nutau} P_{\nutau}(\etaevt) \approx \hat{f}_{\nutau}(\etaevt, \hat{h}_{\nutau})
\end{equation}
\end{linenomath*}
and 
\begin{linenomath*}
\begin{equation}
N_{\nottau} P_{\nottau}(\etaevt) \approx \sum_{\alpha = e, \mu} 
\hat{f}_{\nualpha}(\etaevt, \hat{h}_{\nualpha}),
\end{equation}
\end{linenomath*}
one obtains the tauness
\begin{linenomath*}
\begin{equation}
	P_{\tau}=\frac{\hat{f}_{\nutau}(\etaevt, \hat{h}_{\nutau})}{\sum_{\alpha = e, \mu, \tau} \hat{f}_{\nualpha}(\etaevt, \hat{h}_{\nualpha})}.
	\label{eq:taunessfinal}
\end{equation} 
\end{linenomath*}
Here, $\hat{f}_{\nualpha}(\etaevt, \hat{h}_{\nualpha})$ is the density of $\nualpha$ for the optimal bandwidth $\hat{h}_{\nualpha}$ determined by the \textit{rodeo} algorithm in the region of interest.
Originally developed for unweighted events, we extend the \textit{rodeo} formalism to weighted events according to the procedure in \cite{SAY}: Each of the $n$ simulated events has a weight $w_i$, with $i = {1,...,n}$. We use the effective number of events $n_{\mathrm{Eff}}=(\sum_i w_i)^2 / \sum_i(w_i^2) $, and their effective weight $w_{\mathrm{Eff}}=\sum_i w_i^2 / \sum_i w_i$.

Note that the tauness is always evaluated under certain assumptions for the flux parameters. Computing the tauness for each of the events to originate from a $\nutau$ interaction for the best-fit spectrum given in \cite{HESE7} with a $1/3:1/3:1/3$ flavor composition yields $P_{\tau \, \mathrm{best\,fit}}^{\mathrm{BB}} \approx 75\%$ for ``Big Bird,'' and  $P_{\tau \, \mathrm{best\,fit}}^{\mathrm{DD}} \gtrsim 97\%$ for ``Double Double.'' For ``Double Double,'' the statistics of the generated MC are not sufficient to evaluate the tauness to a higher precision.
The tauness weakly depends on the astrophysical spectral index and decreases by $\sim 1\%$ for a softening of $\gammaa$ by one unit. 

We sample the posterior probability in the flavor composition, obtained by leaving the source flavor composition unconstrained and taking the uncertainties in the neutrino mixing parameters into account. 
When using the best-fit spectra given in \cite{HESE7} but varying the source flavor composition over the entire parameter space (i.e. $\nuratio=a:b:1-a-b$ with $0\leq a,b \leq 1$ and $a+b\leq 1$ at source), and the mixing parameters in the global fit \texttt{NuFit4.1}~\cite{NuFit41} $3\sigma$ allowed range, the tauness is $(97.5_{-0.6}^{+0.3})\%$ for ``Double Double'' and $(76_{-7}^{+5})\%$ for ``Big Bird.''

``Double Double'' is also identified as a candidate tau neutrino event in two complementary analyses using the \textit{double pulse} method to search for tau neutrinos that have been performed while this analysis was ongoing~\cite{ICRCDP1,ICRCDP2}. 
\section{Flavor Composition Analysis}
\label{sec:flavorcomp}
A multi-component maximum likelihood fit is performed on the three topological subsamples using PDFs obtained from MC simulations.
We account for the uncertainty due to limited MC statistics by using a variant of the effective likelihood $\likeSAY$, a generalized Poisson likelihood, presented in \cite{SAY} and employed in \cite{HESE7}.
This joint likelihood is composed of the contributions from the independent subsamples single cascades, double cascades, and tracks  ($\text{SC, DC, and T}$, respectively): 
\begin{equation}
    \likeSAY(\vec{\theta}) =  \prod_{t} \prod_{j} \likeSAY^t \left(\mu_j(\vec{\theta});\sigma_j(\vec{\theta});d_j\right),
\end{equation}
where $\vec{\theta}$ are the model parameters, $j$ are the analysis bins, $\mu_j$ is the expected number of events and the variance in the $j-$th bin with statistical uncertainty $\sigma_j$, $d_j$ is the observed number of events in the $j-$th bin, and $t=(\mathrm{SC, DC, T})$ are the event topologies. Each simulated event $i$ has a weight $w_i$ which depends on the model parameters $\vec{\theta}$. The expected number of events is a product of the effective number of simulated events $n_{\mathrm{Eff}}$ and the effective weight, $w_{\mathrm{Eff}}$ introduced in Section~\ref{sec:tauness}: $\mu=w_{\mathrm{Eff}}n_{\mathrm{Eff}}$.

For all topologies, the contributions from atmospheric and astrophysical neutrinos as well as atmospheric muons are taken into account in the likelihood analysis. 
The conventional atmospheric neutrino component is modeled according to the HKKMS calculation \cite{Honda,RKpi}, the prompt atmospheric neutrino component is modeled following the BERSS \cite{BERSS} (for $\nue, \numu$) and MCEq \cite{AFTG} (for $\nutau$) calculations. MCEq is using the SIBYLL-2.3c \cite{PhysRevD.100.103018} model. The muon component is simulated using MUONGUN \cite{JVS} which samples single muons from templates generated by CORSIKA \cite{Corsika} weighted to the Hillas-Gaisser-H4a cosmic-ray model \cite{H4a} and employing the SIBYLL-2.1 hadronic interaction model \cite{Sibyll2.1} in the shower development.
For the spectrum of the astrophysical neutrino flux $\Phia$, a single power law with a common spectral index $\gammaa$ for all flavors is used,
\begin{linenomath*}
\begin{equation}
    \frac{\dd \Phia}{\dd E} = \sum_{\alpha} \, \phi_{\nualpha} \cdot  \left ( \frac{E}{E_0} \right )^{\gammaa},
	\label{eq:Phiflavor}
\end{equation}
\end{linenomath*}
where $\phi_{\nualpha}$ is the astrophysical normalization of the $\nu+\overline{\nu}$ flux of flavor $\alpha$ at $E_0=100$~TeV.

While for single cascades and tracks, atmospheric contributions pose the main background to the astrophysical signal, the
main background to $\nutau$-induced double cascades arises from misclassified astrophysical $\nu_e$ and $\numu$. The background contributions from atmospheric neutrinos are small (0.2 events in 7.5 years expected), while those from penetrating atmospheric muons and prompt atmospheric $\nutau$ are negligible (0 and 0.04 events in 7.5 years expected, respectively).

The systematic uncertainties are given in Table~\ref{tbl:priors} found in Appendix~C (reproduced from Table V of \cite{HESE7}), and are included in this analysis in the same way as in \cite{HESE7}.
The main systematic uncertainty affecting the double-cascade reconstruction is the anisotropy of the light propagation in the ice \cite{IcePaper,Chirkin:2013lpu}.

While in \cite{HESE7,HESE-xs,HESEBSMICRC,HESEDMICRC}, the total likelihood is maximized assuming flavor equipartition, here we fit the three flavors' fractions $f_{\alpha}$ of the overall astrophysical normalization $\Phia$, $f_{\alpha}=\Phi_{\nu_\alpha}/ \Phia$, with the constraint $f_e +f_{\mu} +f_{\tau} =1$. 
To perform the flavor composition measurement using the multidimensional KDE, the likelihood is modified compared to the analyses in \cite{HESE7}. In the joint likelihood for the three topologies, $\likeSAY = \likeSAY^{\rm SC}  \likeSAY^{\rm T} \likeSAY^{\rm DC}$ \cite{HESE7}, $\likeSAY^{\rm DC}$ is replaced by the extended unbinned likelihood for the double-cascade events, 
\begin{linenomath*}
\begin{equation}
\mathcal{L_{\rm Rodeo}^{\rm DC}} = e^{-\sum_c N_c} \prod_{\mathrm{evt}}\left(\sum_c N_c P_c(\vec \eta_{\mathrm{evt}})\right),
\label{eq:method_b}
\end{equation}
\end{linenomath*}
where $c$ are the flux components used in the fit, $c = \nu_{\text{astro},\alpha}, \nu_{\text{conv}, \alpha}, \nu_{\text{prompt}, \alpha}, \mu_{\text{atm}}$ for the flavors $\alpha=e, \mu, \tau$. $N_c P_c(\etaevt)$ is computed using the \textit{rodeo} algorithm introduced in Section~\ref{sec:tauness}. The aforementioned slight dependence on $\gamma_{\text{astro}}$ is parametrized in the extended double-cascade likelihood $\like^{\rm DC}_{\rm Rodeo}$ by evaluating $N_c P_c(\etaevt)$ as a function of $\gammaa$.
\begin{figure}[t]
\centering
\includegraphics[width=86mm,clip]{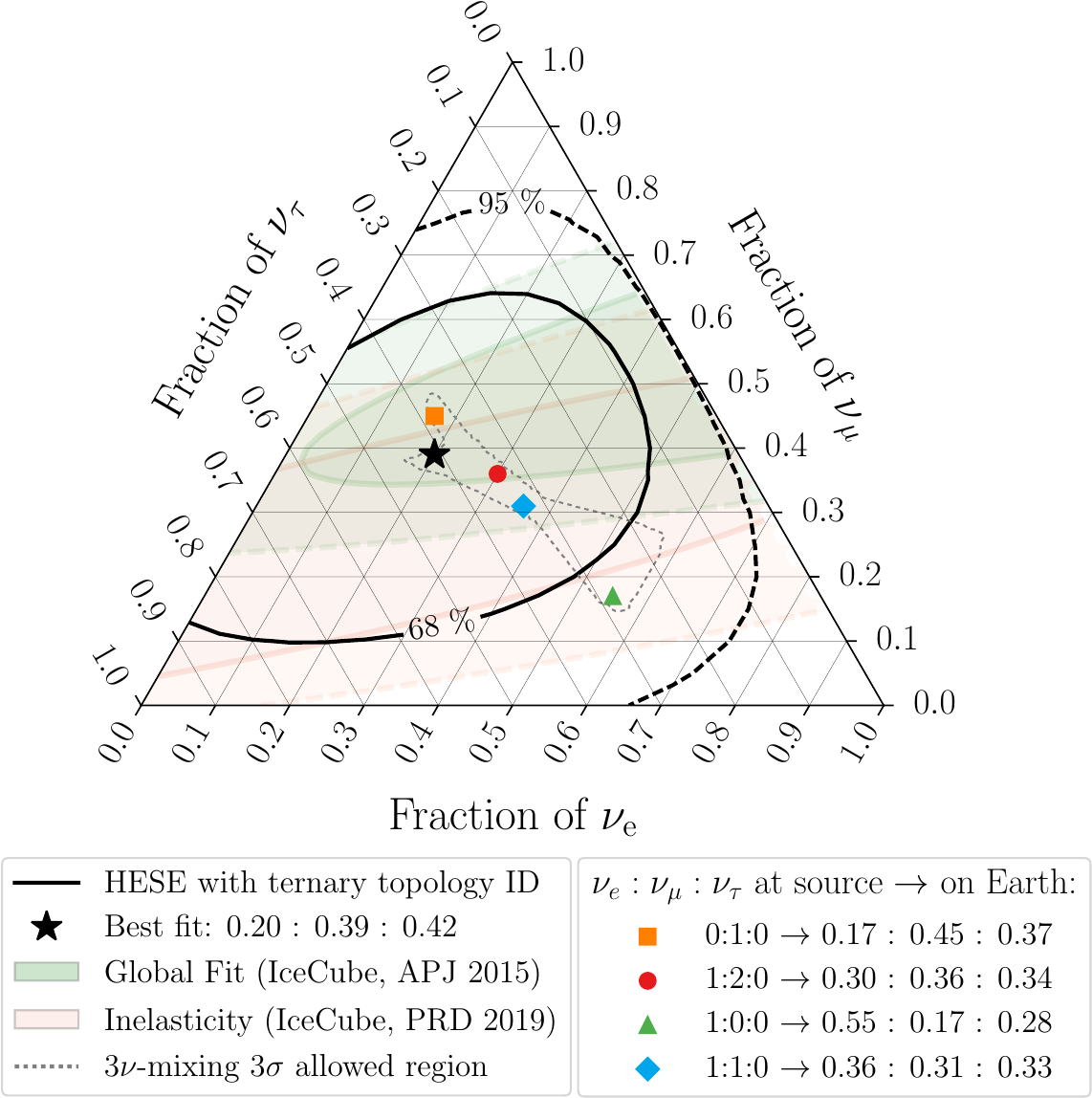}
\vspace{-3mm}
\caption[Measured flavor composition of IceCube HESE events with ternary topology ID and extended multi-dimensional analysis of the double cascades]{Measured flavor composition of IceCube HESE events with ternary topology ID and extended multi-dimensional analysis of the double cascades (black). Contours show the $1\sigma$ and $2\sigma$ confidence intervals assuming Wilks' theorem \cite{Wilks} holds. The shaded regions show previously published results \cite{APJ15,Spencer} without direct sensitivity to the tau neutrino component. Flavor compositions at source and after propagation expected from various astrophysical neutrino production mechanisms (see, e.g., \cite{mubeam-nu}) are marked, and the entire accessible range of flavor compositions assuming standard 3-flavor mixing is shown.}
 \label{fig:KDEflavor}       
\end{figure}

The result of the flavor composition measurement is shown in Figure \ref{fig:KDEflavor}. The fit yields 
\begin{linenomath*}
\begin{equation} \begin{split}
		\frac{\dd \Phia}{\dd E}=&7.4_{-2.1}^{+2.4} \cdot  \left ( \frac{E}{100\text{~TeV}} \right )^{-2.87[-0.20,+0.21]} \\ & \cdot 10^{-18} \cdot  \mathrm{\, GeV^{-1} \, cm}^{-2} \mathrm{\, s}^{-1} \mathrm{\, sr}^{-1},
		\end{split}
\label{eq:KDEbf} 
\end{equation}
\end{linenomath*}
with a best-fit flavor composition of 
\begin{linenomath*}
\begin{equation}
    \nuratio=0.20:0.39:0.42.
\end{equation}
\end{linenomath*}
Comparing this result with previously published results of the flavor composition also shown in Figure~\ref{fig:KDEflavor} clearly shows the advantages of the ternary topological classification. The best-fit point is non-zero in all flavor components for the first time, and the degeneracy between the $\nue$ and $\nutau$ fraction is broken. The small sample size of 60 events in this analysis and the lower sensitivity of the HESE sample to $\numu$ than to $\nue$ and $\nutau$ flavors both lead to an increased uncertainty on the $\numu$ fraction as compared to \cite{APJ15} and~\cite{Spencer}.

The test statistic $\text{TS}= -2 \left( \ln \like(\phi_{\nutau}^0) - \ln \like(\phi_{\nutau}^{\text{b.f.}}) \right)$ compares the likelihood of a fit with a $\nutau$ flux normalization fixed at a value $\phi_{\nutau}^0$ to the free fit where $\phi_{\nutau}$ assumes its best-fit value, $\phi_{\nutau}^{\text{b.f.}}$. 
Evaluated at $\phi_{\nutau}^0=0$ and using Wilks' theorem, it gives the significance at which a vanishing astrophysical tau neutrino flux can be disfavored. 
The test statistic is expected to follow a half-$\chi^2_{k}$ distribution with $k=1$ degree of freedom \cite{halfChi2}. The validity of Wilks' theorem was tested with pseudo-MC trials as described in Appendix~B. The observed test statistic is TS~$=6.5$, which translates to a significance of $2.8\sigma$, or a p-value of 0.005. A one-dimensional scan of the astrophysical $\nutau$ flux normalization is performed with all other components of the fit profiled over. The $1 \sigma$ confidence intervals are defined by TS~$\leq 1$, and the astrophysical tau neutrino flux normalization is measured to 
\begin{linenomath*}
\begin{equation}
	\phi_{\nutau}=3.0_{-1.8}^{+2.2} \cdot 10^{-18} \mathrm{\, GeV^{-1} \, cm}^{-2} \mathrm{\, s}^{-1} \mathrm{\, sr}^{-1}.
	\label{eq:PhiTauresult}
\end{equation}
\end{linenomath*}
This constitutes the first indication for tau neutrinos in the astrophysical neutrino flux. 
\section{Summary and Outlook}
Seven and a half years of HESE events were analyzed with new analysis tools. The previously shown data set was reprocessed with improved detector calibration. A flavor composition measurement was performed using a ternary topological classification directly sensitive to tau neutrinos, which breaks the degeneracy between $\nu_{e}$ and $\nu_{\tau}$ events that is present in a binary classification scheme (into tracks and cascades). This analysis found the first two double cascades, indicative of $\nu_{\tau}$ interactions, with an expectation of 1.5 $\nu_{\tau}$-induced signal events and 0.8 $\nu_{e,\mu}$-induced background events for the best-fit single-power-law spectrum with flavor equipartition \cite{HESE7}. The first event, ``Big Bird,'' has an energy asymmetry at the boundary of the selected interval for double cascades. 
For the second event, ``Double Double,'' the photon arrival pattern is well described with a double-cascade hypothesis, but not with a single-cascade hypothesis. A dedicated \textit{a posteriori} analysis was performed to determine the compatibility of each of the events with a background hypothesis, based on targeted MC. 
The analysis confirms the compatibility of ``Big Bird'' with a single cascade, induced by a $\nu_e$ interaction, at the 25\% level. A ``Big Bird''-like event is $\sim 3$ (15) times more likely to be induced by a $\nu_{\tau}$ than a $\nu_e$ ($\nu_{\mu}$), the result being only weakly dependent on the astrophysical spectral index. ``Double Double'' is $\sim 80$ times more likely to be induced by a $\nu_{\tau}$ than either a $\nu_e$ or a $\nu_{\mu}$. All background interactions have a combined probability of $\sim2\%$, almost independent of the spectral index of the astrophysical neutrino flux. 

Using a novel extended likelihood for double cascades, which allows for the incorporation of a multi-dimensional PDF as evaluated by a kernel density estimator, the flavor composition was measured. The best fit is $\nu_e : \nu_{\mu} : \nu_{\tau} = 0.20:0.39:0.42$, consistent with all previously published results by IceCube \cite{APJ15,Spencer}, as well as with the expectation for astrophysical neutrinos assuming standard 3-flavor mixing. The astrophysical tau neutrino flux is measured to: 
\begin{linenomath*}
\begin{equation} \begin{split}
		\frac{\dd \Phi_{\nutau}}{\dd E}=&3.0_{-1.8}^{+2.2} \left ( \frac{E}{100\text{~TeV}} \right )^{-2.87[-0.20,+0.21]} \\ & \cdot 10^{-18} \cdot  \mathrm{\, GeV^{-1} \, cm}^{-2} \mathrm{\, s}^{-1} \mathrm{\, sr}^{-1}.
		\end{split}
\label{eq:KDEbfnutau} 
\end{equation}
\end{linenomath*}
A zero $\nutau$ flux is disfavored with a significance of $2.8 \sigma$, or, $p = 0.005$.

A limitation of the analysis presented here is the small sample size of 60 events. Merging the HESE selection with the contained cascades event selection \cite{CCscd} is expected to enhance the number of identifiable $\nutau$ events by $\sim 40 \%$ \cite{JSICRC2017}. 
Due to the small effective volume for $\numu$-CC interactions of HESE, the $\numu$ fraction of the astrophysical neutrinos has large uncertainties. Work on updating the joint analysis of multiple event selections \cite{APJ15} is ongoing, where the strongest contribution to constraining the $\numu$ fraction is expected from through-going muons \cite{TGM1, TGMICRC}. 
A few years from now, the IceCube Upgrade \cite{ICRCUpgrade} will greatly improve our knowledge and modeling of the optical properties of the South Pole ice sheet, which the $\nutau$-identification, via the double-cascade method, is sensitive to. The better modeling is expected to lead to a better distinction between single and double cascades around and below the length threshold of 10~m applied in this analysis. The planned IceCube-Gen2 facility \cite{Gen2WP} will provide an order-of-magnitude larger sample of astrophysical neutrinos and enable a precise measurement of their flavor composition, allowing to distinguish between neutrino production mechanisms \cite{Mudamping,mubeam-nu,neutron-nu,charm-nu} with high confidence.  
\\
\begin{acknowledgements}
The IceCube collaboration acknowledges the significant contributions to this manuscript from Carlos Arg\"uelles, Austin Schneider, and Juliana Stachurska. \\
The authors gratefully acknowledge the support from the following agencies and institutions: USA {\textendash} U.S. National Science Foundation-Office of Polar Programs,
U.S. National Science Foundation-Physics Division,
Wisconsin Alumni Research Foundation,
Center for High Throughput Computing (CHTC) at the University of Wisconsin{\textendash}Madison,
Open Science Grid (OSG),
Extreme Science and Engineering Discovery Environment (XSEDE),
Frontera computing project at the Texas Advanced Computing Center,
U.S. Department of Energy-National Energy Research Scientific Computing Center,
Particle astrophysics research computing center at the University of Maryland,
Institute for Cyber-Enabled Research at Michigan State University,
and Astroparticle physics computational facility at Marquette University;
Belgium {\textendash} Funds for Scientific Research (FRS-FNRS and FWO),
FWO Odysseus and Big Science programmes,
and Belgian Federal Science Policy Office (Belspo);
Germany {\textendash} Bundesministerium f{\"u}r Bildung und Forschung (BMBF),
Deutsche Forschungsgemeinschaft (DFG),
Helmholtz Alliance for Astroparticle Physics (HAP),
Initiative and Networking Fund of the Helmholtz Association,
Deutsches Elektronen Synchrotron (DESY),
and High Performance Computing cluster of the RWTH Aachen;
Sweden {\textendash} Swedish Research Council,
Swedish Polar Research Secretariat,
Swedish National Infrastructure for Computing (SNIC),
and Knut and Alice Wallenberg Foundation;
Australia {\textendash} Australian Research Council;
Canada {\textendash} Natural Sciences and Engineering Research Council of Canada,
Calcul Qu{\'e}bec, Compute Ontario, Canada Foundation for Innovation, WestGrid, and Compute Canada;
Denmark {\textendash} Villum Fonden, Danish National Research Foundation (DNRF), Carlsberg Foundation;
New Zealand {\textendash} Marsden Fund;
Japan {\textendash} Japan Society for Promotion of Science (JSPS)
and Institute for Global Prominent Research (IGPR) of Chiba University;
Korea {\textendash} National Research Foundation of Korea (NRF);
Switzerland {\textendash} Swiss National Science Foundation (SNSF);
United Kingdom {\textendash} Department of Physics, University of Oxford.
\end{acknowledgements}
\section*{Appendix A: Targeted MC simulation of the double cascades} 
\label{sec:App_resim}
\begin{table}[thb]
	\centering
	\begin{tabular}{l | c c c }
	%& generated [Millions] & pass $E_{vis}$ \& HESE & Double Cascade class \\
	& generated & pass $E_{vis}$ & Double Cascade \\
	& [Millions] & \& HESE & classification \\
	\midrule
	``Big Bird'' & & & \\
	\hline
	$\nue$ CC & 1.0 & 28\% & 0.7\% \\
	$\nue$ NC & 2.0 & 1\% & 0.05\% \\
	$\nue$ GR & 0.4 & 3\% & 0.1\% \\
	$\numu$ CC & 4.0 & 5\% & 0.02\% \\
	$\numu$ NC & 2.0 & 1\% & 0.05\% \\
	$\nutau$ CC & 2.0 & 19\% & 10.3\% \\
	$\nutau$ NC & 2.0 & 1\% & 0.05\% \\
	\midrule
	``Double Double'' & & & \\
	%``Double  & & & \\
	%Double'' & & & \\
	\hline
	%& generated [Millions] & pass $E_{vis}$ \& HESE & Double Cascade class \\
	$\nue$ CC & 10 & 66\% & 0.77\% \\
	$\nue$ NC & 10 & 5\% & 0.08\% \\
	$\nue$ GR & 2 & 0.3\% & 0.006\% \\
	$\numu$ CC & 40 & 23\% & 0.18\% \\
	$\numu$ NC & 10 & 5\% & 0.08\%\\
	$\nutau$ CC & 10 & 36\% & 1.83\% \\
	$\nutau$ NC & 10 & 5\% & 0.08\% \\
	$\mu$ & $12 \cdot 10^3$ & 0 & 0 \\
	\end{tabular}
	\caption{\modified{Targeted simulation statistics for ``Big Bird" and ``Double Double". The number of total generated events (left), fraction of events passing the HESE selection and the visible energy ($E_{vis}$) requirement (center), and fraction of events classified as double cascades (right) are shown for all classes of simulated interactions.}} \label{tab:resimstats}
\end{table}
The initial, untargeted simulation contains $\sim 7.4 \cdot 10^5$ events in the entire HESE analysis range and thus has insufficient statistics for events similar to the ones observed to calculate the probability for each double cascade to have been induced by a tau neutrino. Targeted MC sets were produced to obtain a large number of MC events with similar properties to the observed double cascade data. Such a simulation is computationally expensive, therefore the targeted MC was restricted to a parameter space around the reconstructed parameters of the observed events, as shown in Table \ref{tab:resim}.

The mapping between true and reconstructed quantities is not straightforward. The interaction vertex in the targeted simulation was restricted to a cylinder with radius 50\,m and height 50\,m, the direction of the incoming neutrino spans $\pm 35 \degree$ in zenith and $\pm 110 \degree$ in azimuth, centered on the reconstructed interaction vertex and direction of the events, respectively. For the zenith and azimuth angles, the resolution depends on the event topology. The azimuth region was chosen to cover a wide range to account for possible contributions from azimuthal regions affected by the ice anisotropy and due to the limited azimuthal resolution for single cascades. The zenith region was restricted more as the zenith resolution is better due to the much closer spacing of DOMs in the vertical direction. For event \#1 simulated as $\numu$ CC interactions, the zenith and azimuth were restricted to $\pm 17\degree$ and $\pm 40\degree$, respectively, to enhance the number of MC events with properties similar to the data, and reflecting the better angular resolution for tracks.
In the case of the primary energy, the mapping depends on the neutrino spectrum and the interaction type, and is only well correlated to the reconstructed deposited energy for $\nue$ CC interactions, as only in this case the neutrino deposits its entire energy in the form of visible energy in the detector. All other interactions have some non-visible energy losses -- final state neutrinos, intrinsically darker hadronic cascades, muons leaving the detector -- such that it is not \textit{a priori} known what primary energy range will significantly contribute to the region around the reconstructed properties of the data events. The primary neutrino energy was restricted to cover the range of energies that can contribute 
to the observed reconstructed energies, % in reconstructed space,
which had to be determined by trial and error \modified{for each simulated interaction type}. True quantities for the energy asymmetry and double cascade length are only defined for $\nutau$ CC interactions. Those properties were therefore left unconstrained during the targeted simulation. 

\modified{Table~\ref{tab:resimstats} lists how many events were simulated for each of the interaction types and what fractions of the simulated events pass the visible energy requirements and HESE selection. These events ($\sim 1\cdot 10^6$ ``Big Bird''-like and $\sim 21 \cdot 10^6$ ``Double Double''-like events) are used in the tau neutrino probability assessment and the flavor composition analysis. For reference, the fraction of events classified as double cascades according to the procedure described in Table~\ref{tab:classification} are also given. } 
\section*{Appendix B: Effect of extended likelihood}
\label{sec:App_rodeo}
\begin{figure}[tb]
\centering
\includegraphics[width=86.5mm,clip]{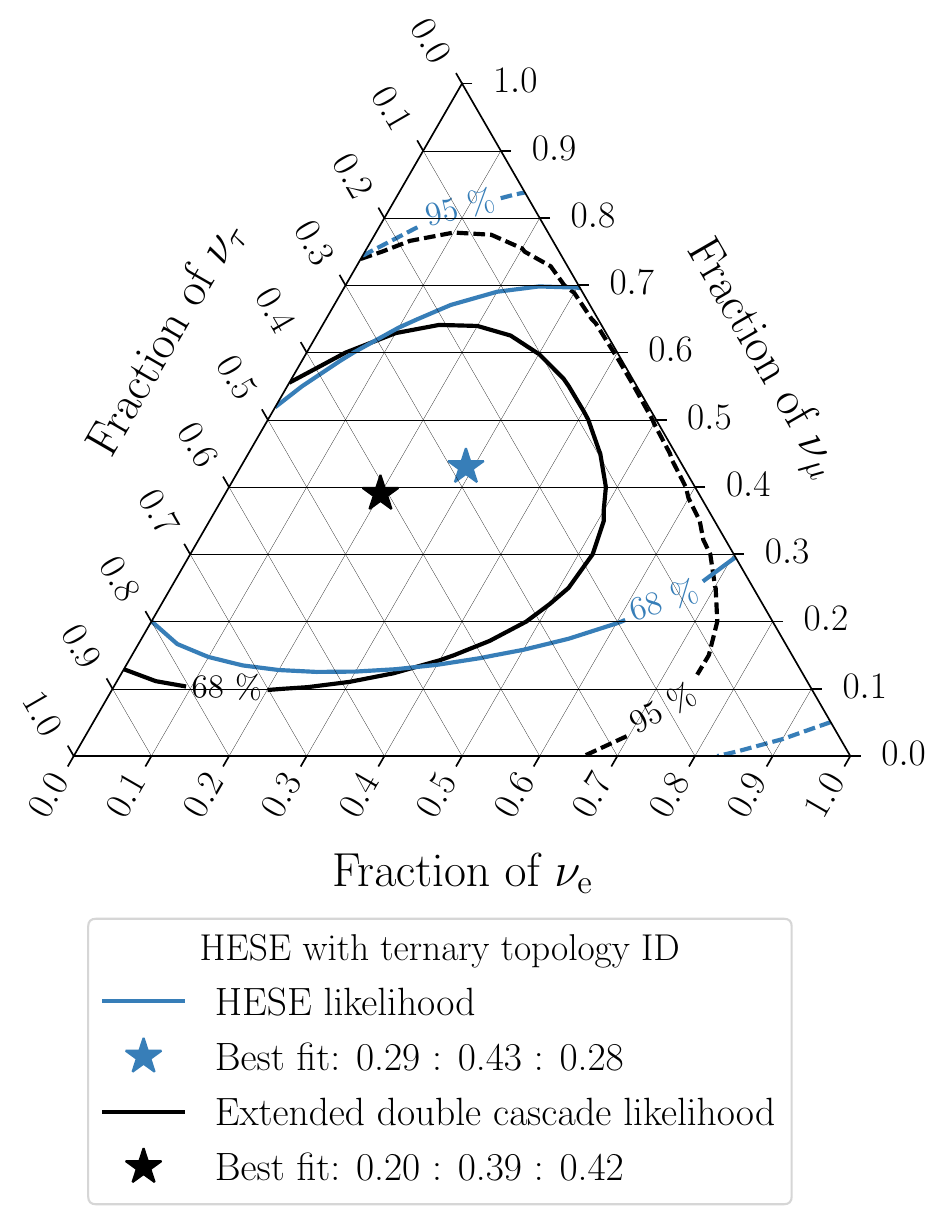}
\caption{Comparison of the flavor composition measurement using the HESE likelihood with ternary topology identification and 2D MC distributions for each topology (blue) with the measurement using the extended likelihood for double cascades as shown in Figure \ref{fig:KDEflavor}.} 
 \label{fig:flavorcomp} %}
\end{figure} 
%\\
Figure \ref{fig:flavorcomp} shows the flavor composition measurement using two-dimensional distributions for all three topologies ($\cos(\theta_z)$ and $\etot$ for single cascades and tracks, $\ldc $ and $\etot$ for double cascades) that are employed in the analyses presented in \cite{HESEBSMICRC,HESE7,HESE-xs,HESEDMICRC}, and the fit using the extended likelihood and targeted simulation for double cascades as shown in Figure \ref{fig:KDEflavor} and used in this analysis.
\begin{figure}[tb]
\centering
\includegraphics[width=86.5mm,clip]{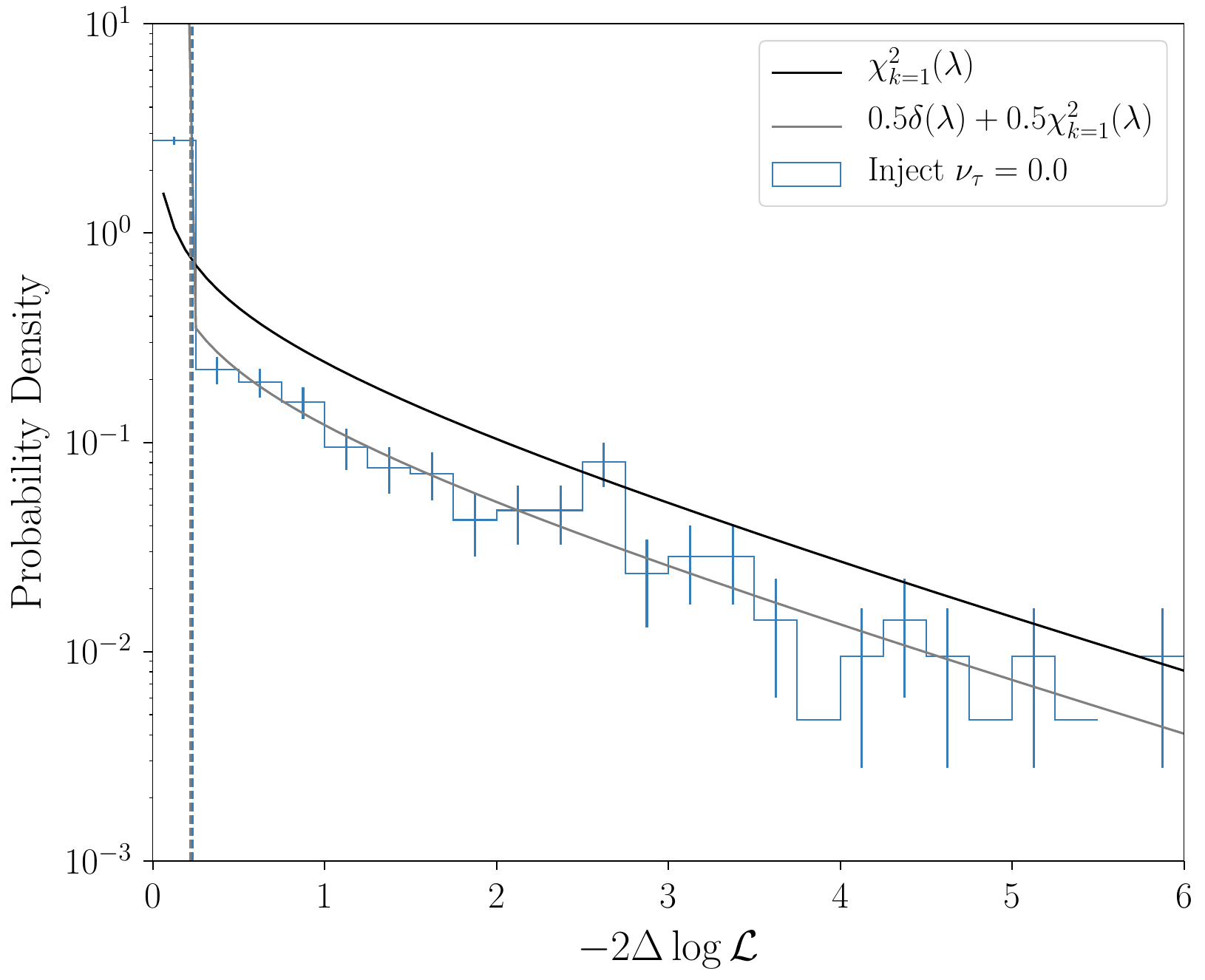}
\caption{Test statistic distribution of the one-dimensional pseudo Monte Carlo trials for an injected vanishing astrophysical tau neutrino flux. The expected half-$\chi^2_{k=1}$ distribution (gray line) is followed. The $\chi^2_{k=1}$ distribution is shown for reference (black line). %, but as expected does not describe the generated distribution of pseudo MC experiments well. 
The vertical lines that mark the events contributing to the 68\% confidence interval (blue), and the expectation from the half-$\chi^2_{k=1}$ distribution (gray) are almost perfectly aligned.} 
 \label{fig:pseudoMC} %}
\end{figure} 

The contours shown in Figure~\ref{fig:KDEflavor} and Figure~\ref{fig:flavorcomp} \modified{\deprecated{here }}were obtained assuming Wilks' theorem holds. The validity of Wilks' theorem was tested for the untargeted MC, by generating pseudo-MC trails. For a one-dimensional fit as used for the astrophysical $\nutau$ flux normalization measurement, Wilks' theorem holds over the entire available parameter space for the untargeted MC. The generated pseudo-MC trials are distributed according to a half-$\chi^2_{k=1}$ distribution, as can be seen in Figure~\ref{fig:pseudoMC}. For the two-dimensional fit used for the flavor composition, Wilks' theorem holds within the flavor triangle and gives a conservative result at the boundary where one of the contributions vanishes. 
%
%\blue{
\section*{Appendix C: Analysis parameters and systematic uncertainties in the HESE analyses}
\label{sec:App_syst}
\begin{table*}[thb]
	\centering
	\begin{tabular}{l c r c r c r}
		\toprule
		Parameter & $\, $ $\, $ & Constraint & $\, $ $\, $ & Range &$\, $ $\, $  & Description \\
		\hline
		\midrule
		\multicolumn{1}{l }{\textbf{Astrophysical neutrino flux:}} & & & &&&\\
		$\Phia$ & & - & &  $[0,\infty)$ & &  Normalization scale\\
		$\gamma$ & & - & &   $(-\infty,\infty)$ & & Spectral index\\
		$f_{\alpha}$ & & $\sum_{\alpha} f_{\alpha}=1$ & &   $[0,1]$ & & Relative flavor contribution\\
		\\
		%\hline
		%\rule{0pt}{4mm}\ignorespaces
		\midrule
		\multicolumn{1}{l }{\textbf{Atmospheric neutrino flux:}} & & & &&& \\
		$\phi_{\mathrm{conv}}$ & & $1.0\pm0.4$ & & $[0, \infty)$ & & Conventional normalization scale\\
		$\phi_{\mathrm{prompt}}$ & & - & &  $[0, \infty)$ & & Prompt normalization scale\\
		$R_{K/\pi}$ & & $1.0\pm0.1$ &  & $[0, \infty)$ & & Kaon-pion ratio correction\\
		$2 \nu /(\nu + \overline{\nu})_{\mathrm{atmo}}$  & & $1.0\pm0.1$ & & $[0,2]$ & & Neutrino-antineutrino ratio correction\\
		\\
		%\hline
		%\rule{0pt}{4mm} %\ignorespaces
		\midrule
		\multicolumn{1}{l }{\textbf{Cosmic-ray flux:}} & & & &&&\\
		$\Delta \gamma_{\mathrm{CR}}$ & & $0.0\pm 0.05$ & & $(-\infty,\infty)$ & & Cosmic ray spectral index modification\\
		$\Phi_{\mu}$ & & $1.0\pm 0.5$ & & $[0,\infty)$ & & Muon normalization scale\\
		\\
		%\hline
		%\rule{0pt}{4mm}\ignorespaces
		\midrule
		\multicolumn{1}{l }{\textbf{Detector:}} & & & &&&\\
		$\epsilon_{\mathrm{DOM}}$ & & $0.99 \pm 0.1$ & & $[0.80, 1.25]$ & & DOM efficiency\\
		$\epsilon_{\mathrm{head-on}}$ & & $0.0 \pm 0.5$ & & $[-3.82, 2.18]$ & & DOM angular response\\
		$a_s$ & & $1.0 \pm 0.2$ & & $[0.0, 2.0]$ & & Ice anisotropy scale\\
		\bottomrule
	\end{tabular}
	\caption{Analysis model parameters for the single power-law astrophysical model. Constraints for analysis parameters used in the analysis are shown.
		%These constraints on the parameters are either uniform or Gaussian.
		The mean and standard deviation are given for Gaussian constraints, while constraint-free parameters are denoted with ``-''.
		Bounds are given for all parameters. Note that the $f_{\alpha}$ parameters are unique to the analysis presented here.}\label{tbl:priors}
\end{table*}
The analysis model parameters are given in Table~\ref{tbl:priors}, modified from \cite{HESE7}. The atmospheric fluxes need to be carefully modeled. This is done via the parameters: $\phi_{\mathrm{conv}}$ scaling the overall conventional atmospheric neutrino flux normalization, $\phi_{\mathrm{prompt}}$ scaling the overall \\ prompt atmospheric neutrino flux normalization, $R_{K/\pi}$ scaling the kaon-to-pion ratio, $2 \nu /(\nu + \overline{\nu})_{\mathrm{atmo}}$ providing the neutrino-to-antineutrino ratio, $\Delta \gamma_{\mathrm{CR}}$ accounting for modifications in the cosmic ray spectral index, and $\Phi_{\mu}$ scaling the atmospheric muon flux normalization.
The light yield of an optical module is affected by its overall efficiency and the propagation of photons through the ice to reach the module. Uncertainties in the former are parametrized by the DOM efficiency parameter, $\epsilon_{\mathrm{DOM}}$, which describes changes in the total efficiency of the DOMs. Uncertainties on the ice properties are parameterized with the $\epsilon_{\mathrm{head-on}}$ parameter, which modifies the angular response of the DOM and depends on local ice properties of the refrozen ice surrounding the DOMs, and $a_s$ describing an azimuthal anisotropy of the photon propagation. Events have been simulated using variations of all of these parameters. With the exception of the anisotropy scale $a_s$, the parameter uncertainties affect all topological classes of events in the same way, their effect and treatment is discussed in detail in~\cite{HESE7}.

%\section{Anisotropy of photon propagation in ice:} 
The anisotropic photon propagation in the ice can affect the classification of events and needs careful attention. The ice model used in this analysis is called \textit{Spice3.2}, and contains the South Pole ice sheet's optical properties (scattering and absorption coefficients) at each point in the detector and for each direction of photon propagation. 
Measurements with in-situ IceCube calibration LEDs \cite{IcePaper} have shown that the ice is not isotropic \cite{Chirkin:2013lpu}, i.e., the propagation of a photon depends on its  direction. The anisotropy can be modeled as a sinusoidal modulation of the scattering coefficients in azimuth and zenith. Along the glacial flow direction, scattering is reduced by $-10\%$ while perpendicular to the glacial flow it is enhanced by $+5\%$. Less scattering leads to photons traveling on straighter paths through the ice, which in turn leads to a cascade being elongated when aligned with the glacial flow. More scattering leads to photons traveling on more random paths, thus a cascade becomes compressed when its direction is perpendicular to the glacial flow. If uncorrected, this effect leads to a larger misclassification of true single cascades as double cascades due to their elongation along the glacial flow, and of true double cascades as single cascades due to their compression perpendicular to the glacial flow. 
The event reconstruction algorithm uses look-up tables and compares the received to the expected photon counts per receiving DOM. The look-up tables assume isotropy, and adding dimensions to incorporate the anisotropy would make the photon tables too large and their production as well as each event reconstruction computationally too expensive. As the expected light yield is looked up for each receiving DOM, a simple trick can be used to approximately correct for the anisotropy: the distance between source and receiving DOM is shifted and the expected light yield looked up for the effective distance which contains the effect of enhanced or inhibited photon propagation due to the anisotropy~\cite{MarcelPhD}. This first-order correction of the anisotropy of the photon propagation has been verified as sufficient, as the misclassification fraction of true single cascades as double cascades is now constant across the full azimuth range. To model uncertainties in the anisotropy scale, the scale parameter $a_s$ is used. The length bias is defined as the mean difference of reconstructed lengths when the anisotropy is corrected for and when it is not, and is a function of the reconstructed zenith and azimuth. Under the assumption that the length bias scales linearly with the anisotropy scale parameter, the anisotropy scale can be fit from data. As the length bias is a function of reconstructed observables, the systematic uncertainties on the reconstructed length due to the anisotropy can be tested per event. Note that none of the classified double cascades is in a phase space greatly affected by the anisotropy. The uncertainty on the anisotropy scale in this analysis is 20\%. As the direction of the anisotropy is known with sub-degree precision, no uncertainty is assumed on it.

%\end{document}
\bibliographystyle{apsrev4-1}
\bibliography{bibfile}

\end{document}